\RequirePackage{fix-cm}
\documentclass[smallextended]{svjour3}       
\smartqed  
\usepackage{graphicx}
\usepackage{amssymb}
\begin{document}

\title{Efficiency of Open Quantum Walk implementation of Dissipative Quantum computing algorithms}

\titlerunning{Efficiency of Open Quantum Walk implementation...}        

\author{Ilya Sinayskiy         \and
        Francesco Petruccione 
}


\institute{I. Sinayskiy \at
              NITheP and School of Chemistry and Physics,\\
              University of KwaZulu-Natal, Westville, Durban, South Africa\\
              Tel.: +27-31-260-8133\\
              Fax: +27-31-260-8090\\
              \email{sinayskiy@ukzn.ac.za}           
           \and
           F. Petruccione \at
              NITheP and School of Chemistry and Physics,\\
              University of KwaZulu-Natal, Westville, Durban, South Africa\\
              Tel.: +27-31-260-2770\\
              Fax: +27-31-260-8090\\
              \email{petruccione@ukzn.ac.za} 
}

\date{Received: date / Accepted: date}

\maketitle

\begin{abstract}
An open quantum walk formalism for dissipative quantum computing is presented. The approach is illustrated with the examples of the Toffoli gate and the Quantum Fourier Transform for 3 and 4 qubits. It is shown that the algorithms based on the open quantum walk formalism are more efficient than the canonical dissipative quantum computing approach. In particular, the open quantum walks can be designed to converge faster to the desired steady state and to increase the probability of detection of the outcome of the computation.
\keywords{open quantum walk \and dissipative quantum computing \and quantum Fourier transform}
\PACS{03.65.Yz  \and 05.40.Fb \and 02.50.Ga}
\end{abstract}

\section{Introduction}
The realistic description of any quantum system includes the unavoidable effect of the interaction with the environment \cite{BP}. Such open quantum systems are characterized by the presence of dissipation and decoherence. For many applications, the influence of both phenomena on the reduced systems needs to be eliminated or at least minimized. However, it was shown recently that the interaction with the environment not only can create complex entangled states \cite{ent1,ent2,ent3,ent4,ent5,ent6}, but also allows for universal quantum computation \cite{dqc}.

One of the well established approaches to formulate quantum algorithms is the language of quantum walks \cite{qw1,qw2}. Both, continuous and discrete-time quantum walks can perform universal quantum computation \cite{qwqc1,qwqc2}. Usually, taking into account the decoherence and dissipation in a unitary quantum walk reduces its applicability for quantum computation \cite{ken2} (although, in 
very small amounts decoherence has been found to be useful \cite{ken1}).

Recently, a framework for discrete time open quantum walks on graphs was proposed \cite{longAttal}, which is based upon an exclusively dissipative dynamics. This framework is inspired by a specific discrete time implementation of the Kraus representation of CP-maps on graphs. In continuous time and more general setting Whitfield et al. \cite{QSW} introduced quantum stochastic walks to study the transition from quantum walk to classical random walk. In this paper the flexibility and the strength of the open quantum walk formalism \cite{longAttal} will be demonstrated  by implementing algorithms for dissipative quantum computing. With the example of the Toffoli gate and the Quantum Fourier Transform with 3 and 4 qubits we will show that the open quantum walk implementation of the corresponding algorithms outperforms the original dissipative quantum computing model \cite{dqc}.

In section 2 we briefly summarize the formalism of open quantum walks. In section 3 we review the dissipative quantum computing model and show how to implement an arbitrary simple unitary gate as well as the Toffoli gate. In section 4 with the help of the Quantum Fourier Transform for 3 and 4 qubits we demonstrate that the open quantum walk approach to quantum computing allows for the implementation of more involved quantum algorithms. In section 5 we conclude and present an outlook on future work.

\section{General construction of Open Quantum Walks}

Open Quantum Walks are defined on graphs with a finite or countable number of vertices \cite{longAttal}. The dynamics of the walker will be described in the Hilbert space given by the tensor product $\cal{H}\otimes\cal{K}$. $\cal{H}$ denotes the Hilbert space of the internal degrees of freedom of the walker. For example, in the case of a spin $1/2$ walker the Hilbert space $\mathcal{H}$ is $\mathcal{H}=\mathbb{C}^{2}$. The graph on which the walk is performed is decribed by a set of vertices $\mathcal{V}$. The Hilbert space $\mathcal{K}=\mathbb{C}^\mathcal{V}$ has as many basis vectors, as number of vertices in $\mathcal{V}$. For an infinite number of vertices we consider $\mathcal{K}$ to be any separable Hilbert space with orthonormal basis $(|i\rangle)_{i \in \mathcal{V}}$.

For each edge $(i,j)$ of the graph we introduce a bounded operator $B_j^i\in \mathcal{H}$ which will play the role of a generalized quantum coin. The operator $B_j^i$ describes a transformation in the internal degree of freedom of the walker while ``jumping" from node $j$ to node $i$. To ensure conservation of probability and positivity we enforce the condition,
\begin{equation}
\sum_i B_j^{i\dag}B_j^i=I.
\end{equation}
This condition guarantees that the local map $\mathcal{M}_j(\rho)$ defined at each vertex $j$,
\begin{equation}
\mathcal{M}_j(\rho)=\sum_i B_j^i\rho B_j^{i\dag},
\end{equation}
is completely positive and trace preserving. The CP-map $\mathcal{M}_j$ is defined on the Hilbert space $\mathcal{H}$. In order to extend $\mathcal{M}_j$ to the Hilbert space of the total system, i.e. $\mathcal{H}\otimes\mathcal{K}$, we dilate the generalized quantum coin operation $B_i^j$ with the transition on the graph in the following way, $M_j^i=B_j^i\otimes |i\rangle\langle j|$. It is easy to see, that if the basis vectors $|i\rangle$ are orthonormal basis vectors then $M_j^i$ satisfies the following condition,
\begin{equation}
\sum_{i,j} M_j^{i\dag}M_j^i=I.
\end{equation}
The above equality allow us to define a trace preserving and CP- map $\mathcal{M}$ on the Hilbert space of the total system $\mathcal{H}\otimes\mathcal{K}$, as
\begin{equation}
\mathcal{M}(\rho)=\sum_{i,j} M_j^i\rho M_j^{i\dag}.
\end{equation}
With this choice of operators $M_i^j$ the map $\mathcal{M}$ conserves the structure of the density operators of the following form, 
\begin{equation}
\rho=\sum_i \rho_i\otimes |i\rangle\langle i|, 
\end{equation}
with $\sum_i \mathrm{Tr}(\rho_i)=1$.
In fact, one sees immediately that,
\begin{equation}
\mathcal{M}\left(\sum_i \rho_i\otimes |i\rangle\langle i|\right)=\sum_i\left(\sum_j B_j^i\rho_j B_j^{i\dag}\right)\otimes |i\rangle\langle i|.
\end{equation}
The map $\mathcal{M}$ acting on density matrices of the form $\rho=\sum_i \rho_i\otimes |i\rangle\langle i|$  defines the Open Quantum Walk.

One should understand that within this formulation of the Open Quantum Walk the transition between nodes $i$ and $j$ of the graph are driven purely by the dissipative interaction with a common bath between this two nodes. In this sense the transitions between nodes are environment mediated. The direct transition due to unitary evolution is prohibited. In a corresponding microscopic system-environment model an appropriate total Hamiltonian guaranties that during each step of the walk the ``walker" interacts with the Markovian environment common to the nodes involved in the step. The system-bath interaction is engineered in such a way that during the transition from the node $i$ to the node $j$ a quantum coin $(B_i^j)$ is applied to the internal degree of freedom of the ``walker". From this point the transition operator from the node $i$ to the node $j$ is proportional to $B_i^j\otimes|j\rangle\langle i|$ so that the probability of the ``walker" to jump will depend on the state of the internal degree of freedom and the interaction a with local Markovian environment. A full microscopic derivation of an open quantum walk from a physical Hamiltonian of a total system is beyond the scope of the present paper and will be presented elsewhere \cite{MD}.

\begin{figure}
\begin{center}
\includegraphics[width=0.5\textwidth]{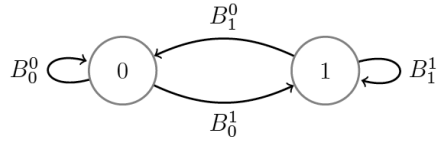}
\caption{A schematic representation of an open quantum walk on a 2-node graph. The operators $B_i^j$ $(i,j=0,1)$ represent the transition operators of the walk.}
\end{center}
\label{figU}       
\end{figure}

As an example of open quantum walk let us consider the simplest case of a walk on a 2-node graph (see Fig. 1). In this case the transition operators $B_i^j$ $(i,j=0,1)$ satisfy:
\begin{equation}
B_0^{0\dag}B_0^0+B_0^{1\dag}B_0^1=I,\quad B_1^{1\dag}B_1^1+B_1^{0\dag}B_1^0=I.
\end{equation}
The state of the walker $\rho^{[n]}$ after $n$ steps is given by,
\begin{equation}
\rho^{[n]}=\rho_0^{[n]}\otimes|0\rangle\langle 0|+\rho_1^{[n]}\otimes|1\rangle\langle 1|,
\label{2node}
\end{equation}
where the particular form of the $\rho_i^{[n]}$ $(i=0,1)$ is found by recursion,
\begin{eqnarray}
\rho_0^{[n]}=B_0^0\rho_0^{[n-1]}B_0^{0\dag}+B_1^0\rho_1^{[n-1]}B_1^{0\dag},\\\nonumber
\rho_1^{[n]}=B_0^1\rho_0^{[n-1]}B_0^{1\dag}+B_1^1\rho_1^{[n-1]}B_1^{1\dag}.
\end{eqnarray}

\section{Implementation of the arbitrary unitary operation and the Toffoli gate}

Recently, Verstraete \textit{et al.} \cite{dqc} suggested a dissipative model of quantum computing, capable of performing universal quantum computation.
The dissipative quantum computing setup consists of a linear chain of time registers. Initially, the system is in a time register labeled by $0$. The result of the computation is measured in the last time register labeled by $T$. Neighboring time registers are coupled to local baths. When the system reaches its unique steady state the result of the planned quantum computational task is the state of the time register $T$. In particular, for a quantum circuit given by the set of unitary operators $\{ U_t\}_{t=1}^T$ the final state of the system is given by $|\psi_T\rangle=U_TU_{T-1}\ldots U_2U_1|\psi_0\rangle$. To reach the final state $|\psi_T\rangle$ one evolves the system with the help of the master equation,
\begin{equation}
\frac{d}{dt} \rho=\sum_k L_k \rho L_k^\dagger - \frac{1}{2} \{ L_k^\dagger L_k, \rho \}_+,
\end{equation}
where jump operators $L_k$ are given by $L_i = |0\rangle_i \langle 1| \otimes |0\rangle_t \langle 0|$ and $L_t=U_t\otimes|t\rangle\langle t+1|+U_t^\dag\otimes|t+1\rangle\langle t|$. Verstraete \textit{et al.} \cite{dqc} have shown that in this case the total system converges to a unique steady state, namely,
\begin{equation}
\rho=\frac{1}{T+1}\sum_t|\psi_t\rangle\langle \psi_t|\otimes|t\rangle\langle t|.
\end{equation}
It is clear, that the probability of successful detection of the result of the quantum computation $|\psi_T\rangle$ is given by $1/(T+1)$.

Using the formalism of open quantum walks one can perform dissipative quantum computations with higher efficiency. In order to demonstrate this fact we consider in the following the open quantum walk implementation of the simple unitary operation and the Toffoli gate.

We start by showing how to implement a simple gate given by the unitary operator $U$. To achieve  this it is sufficient to consider a 2-node graph (see Fig. 1). By choosing the following form of transition operators, $B_0^0=\sqrt{\lambda}I$, $B_1^1=\sqrt{\omega}I$, $B_0^1=\sqrt{\omega}U$ and $B_1^0=\sqrt{\lambda}U^\dag$ the OQW shown in Fig. 1 will implement the single gate $U$. If the initial state of the system $|\psi_0\rangle$ is prepared in the node $0$, then after performing the open quantum walk the system reaches the steady state $\rho_{SS}=\lambda|\psi_0\rangle\langle \psi_0|\otimes|0\rangle\langle 0|+\omega U|\psi_0\rangle\langle \psi_0|U^\dag\otimes|1\rangle\langle 1|$. The positive constants $\omega$ and $\lambda$ satisfy $\lambda+\omega=1$. The result of the gate application can be detected in node $1$ with probability $\omega$. 

The physical meaning of the parameters $\omega$ and $\lambda$ can be understood from the underlying microscopic model of the system \cite{MD}. For a ``walker" coupled to bosonic Markovian baths we expect the parameters $\omega$ and $\lambda$ to scale linearly with the mean number $n$ of thermal bosons (photon or phonons) corresponding to the frequency of transition in the common environment which mediates transitions between nodes,
\begin{equation}
\omega\sim\gamma(n+1)\, \textrm{and} \,  \lambda\sim\gamma n,
\end{equation} 
where $\gamma$ is a coefficient of the spontaneous emission. From this point of view the steady state of the ``walker" on the 2 node graph will always have the form (see Eq. (\ref{2node})),
\begin{equation}
\rho_{SS}=\rho^{[0]}_{SS}\otimes|0\rangle\langle 0|+\rho^{[1]}_{SS}\otimes|1\rangle\langle 1|.
\end{equation}
If one takes $B_i^1\sim\sqrt{\omega}$ and $B_i^0\sim\sqrt{\lambda}$ for $i=(0,1)$, then $\mathrm{Tr}[\rho^{[0]}_{SS}]\sim n$ and $\mathrm{Tr}[\rho^{[1]}_{SS}]\sim (n+1)$.
It is clear that there are two limiting cases, first $\omega=\lambda$ in the very high temperature limit $(T_{\mathrm{Bath}}=\infty)$ and second  $\omega=1, \lambda=0$ in the zero temperature case $(T_{\mathrm{Bath}}=0)$.
\begin{figure}
\begin{center}
\includegraphics[width=0.85\textwidth]{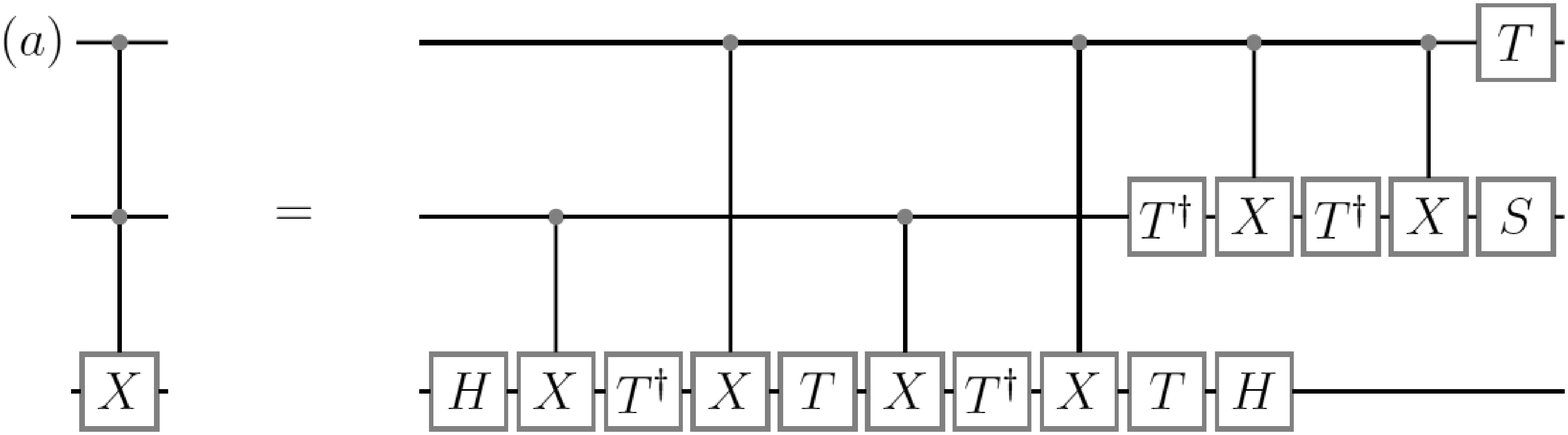}\\
\includegraphics[width=0.95\textwidth]{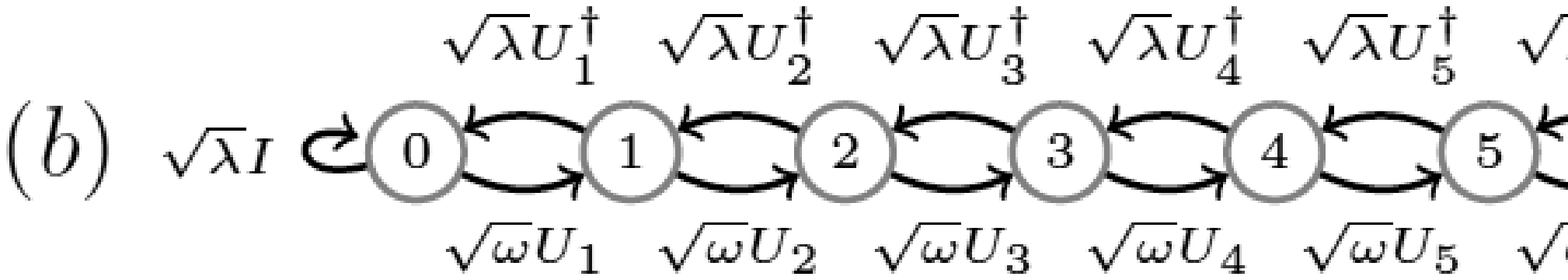}\\
\includegraphics[width=0.46\textwidth]{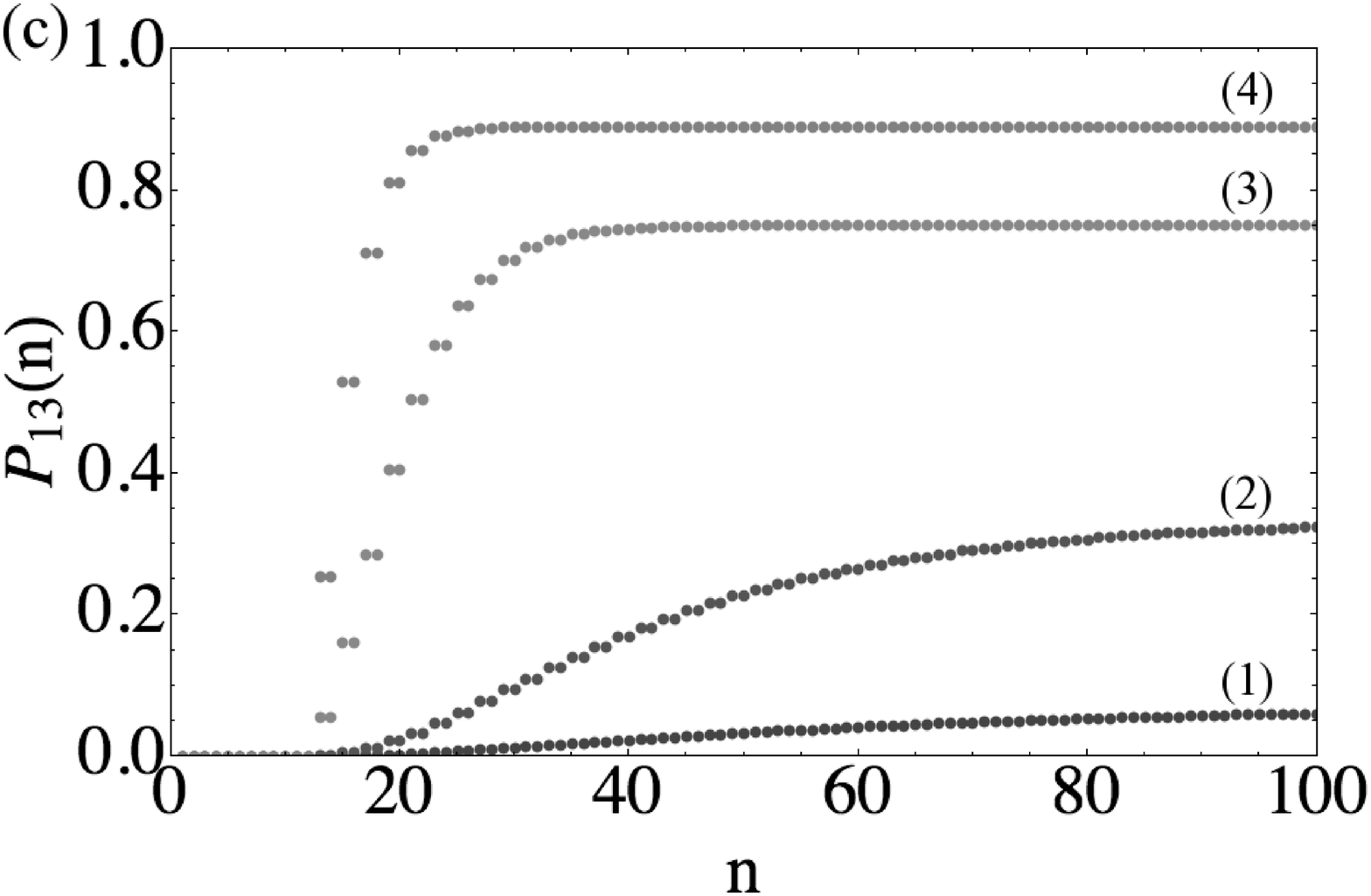}
\includegraphics[width=0.52\textwidth]{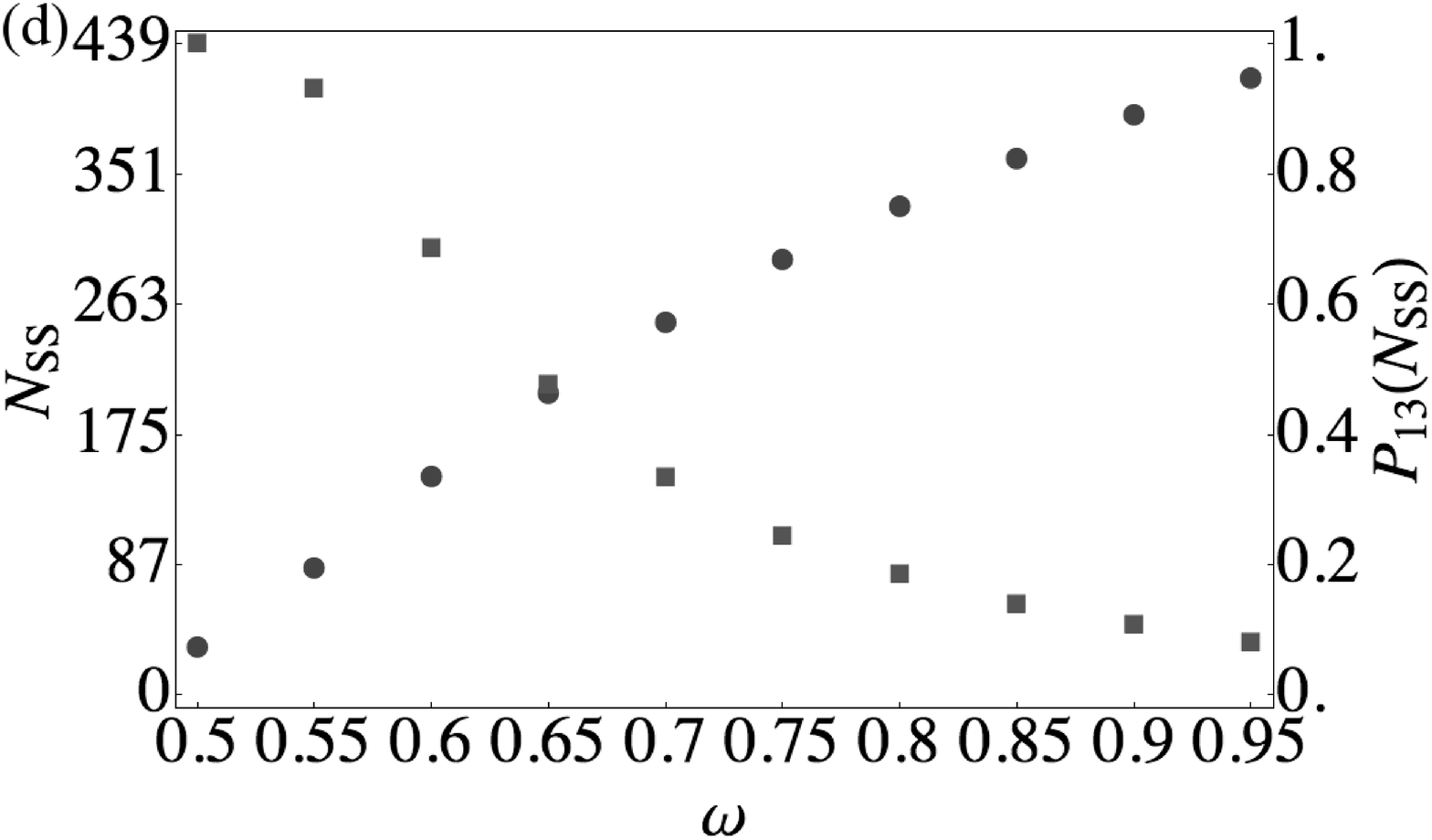}
\caption{Quantum circuit, corresponding open quantum walk diagram and efficiency of the Toffoli gate. Fig. 2a depicts the circuit implementation of the Toffoli gate. The corresponding open quantum walk diagram is shown in Fig. 2b. Fig. 2c shows the dynamics of the detection probability in the final node $13$ as function of the number of steps of the OQW. Curves (c1) to (c4) correspond to different values of the parameter $\omega=0.5, 0.6, 0.8, 0.9$, respectively. Fig. 2d shows the number of steps needed  to reach the steady state (squares) and the probability of detection of the successful implementation of the gate (circles) as function of the parameter $\omega$. The number of steps to reach a steady states is simulated with $10^{-7}$ accuracy.}
\end{center}
\label{figU2}       
\end{figure}

Next we analyze the OQW implementation of the Toffoli gate \cite{nc}. Using single qubits and CNOT-gates the Toffoli gate can be realized in a circuit shown in Fig. 2a. The single qubits gates $S$, $T$, $X$ and $H$ are given by $S=|0\rangle\langle 0|+e^{i\pi/2}|1\rangle\langle 1|$,  $T=|0\rangle\langle 0|+e^{i\pi/4}|1\rangle\langle 1|$,  $X=|0\rangle\langle 1|+|1\rangle\langle 0|$ and  the Hadamard gate $H=\left(|0\rangle\langle 0|-|1\rangle\langle 1|+X\right)/\sqrt{2}$. To implement the Toffoli gate we need to implement 13 unitary operators. In the language of dissipative quantum computing this means that we need $13+1$ time-registers ($T=13$). The corresponding open quantum walk scheme is shown in Fig 2b. In this case each node of the graph corresponds to a 3-qubit Hilbert space and each step of the walk corresponds to a transition of all three qubits. 
The set of unitary operators $U_1,U_2,...,U_{13}$ corresponds to unitaries in the circuit. For example, the unitary operator $U_6$ is given by 
\begin{equation}
U_6=I_2\otimes |0\rangle\langle 0|\otimes I_2 + I_2\otimes |1\rangle\langle 1|\otimes X.
\end{equation}
In the Figs. 2c and 2d we analyze the efficiency of the OQW implementation of the Toffoli gate as a function of the parameter of the walk, i.e., $\omega$. Fig. 2c shows the dependence of the detection probability in the last node labeled by $13$ as function of the number of steps of the open quantum walk for different values of the parameter $\omega$.  Curve (1) of Fig. 2c corresponds to $\omega=0.5$, which is the efficiency of the conventional dissipative quantum computing scheme. The formalism of open quantum walks allows to choose other values for $\omega$. In particular, for higher values of $\omega=0.6, 0.8, 0.9$, the open quantum walk shows a higher efficiency of computation. Fig. 2d analyzes the number of steps needed to reach the steady state and the probability of detection of the result of the computation in the steady state as a function of the parameter $\omega$. From Fig. 2d it is clear that the number of steps needed to reach a steady state decreases with increasing parameter $\omega$. 

The above result has a straightforward interpretation from the open quantum system dynamics point of view. Obviously, the ground state of the total system ``walker+network" is the pure state $|\psi_G\rangle=U_{13}U_{12}\ldots U_2U_1|\psi_0\rangle\otimes|13\rangle$, where $|\psi_0\rangle$ is the 3-qubit input state of the Toffoli gate and $|13\rangle$ labels the 13th node of the graph in Fig. 2b. The steady state of the open walk converges to this pure state only in the case of $\omega=1$ and $\lambda=0$ which corresponds to zero temperature of all local environments. In all other cases the steady state will be given by the density matrix $\rho_{SS}=\sum_{i=0}^{13} p_i |\psi_i\rangle\langle \psi_i|\otimes|i\rangle\langle i|$, where $|\psi_i\rangle=U_iU_{i-1}\ldots U_1|\psi_0\rangle$. In the case when $\omega=\lambda=1/2$, which corresponds to the conventional DQC scheme, all probabilities $p_i=1/(T+1)$, where $T=13$. The probability to find the "walker" in the ground state increases with decreasing temperatures of the local environments, which in turn corresponds to increasing the parameter $\omega$. 
In the explicit implementation of the quantum algorithm the parameter $\omega$ determines the probability of forward propagation. In a similar way it is also obvious from Fig. 2d that with increasing parameter $\omega$  the probability of detection of the result of the computation in node $13$ increases.

\section{Three and four qubit quantum Fourier transform}

\begin{figure}
\begin{center}
\includegraphics[width=0.85\textwidth]{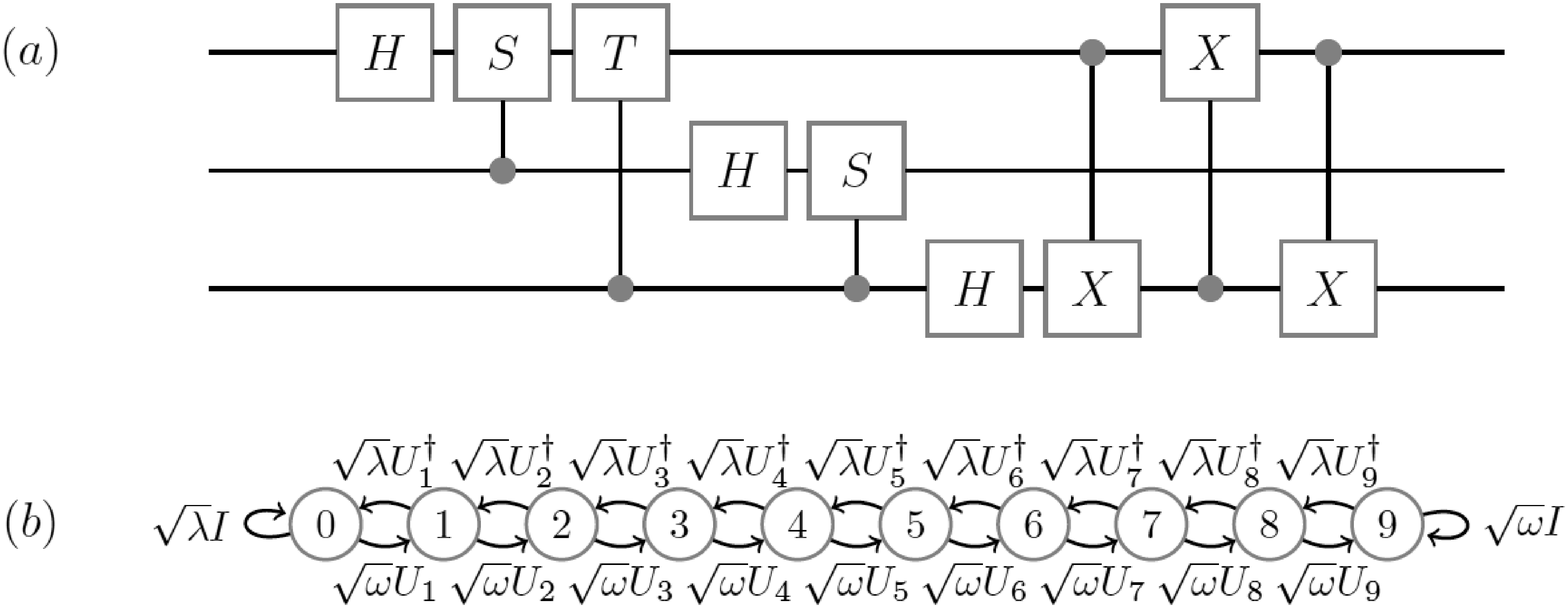}\\
\includegraphics[width=0.46\textwidth]{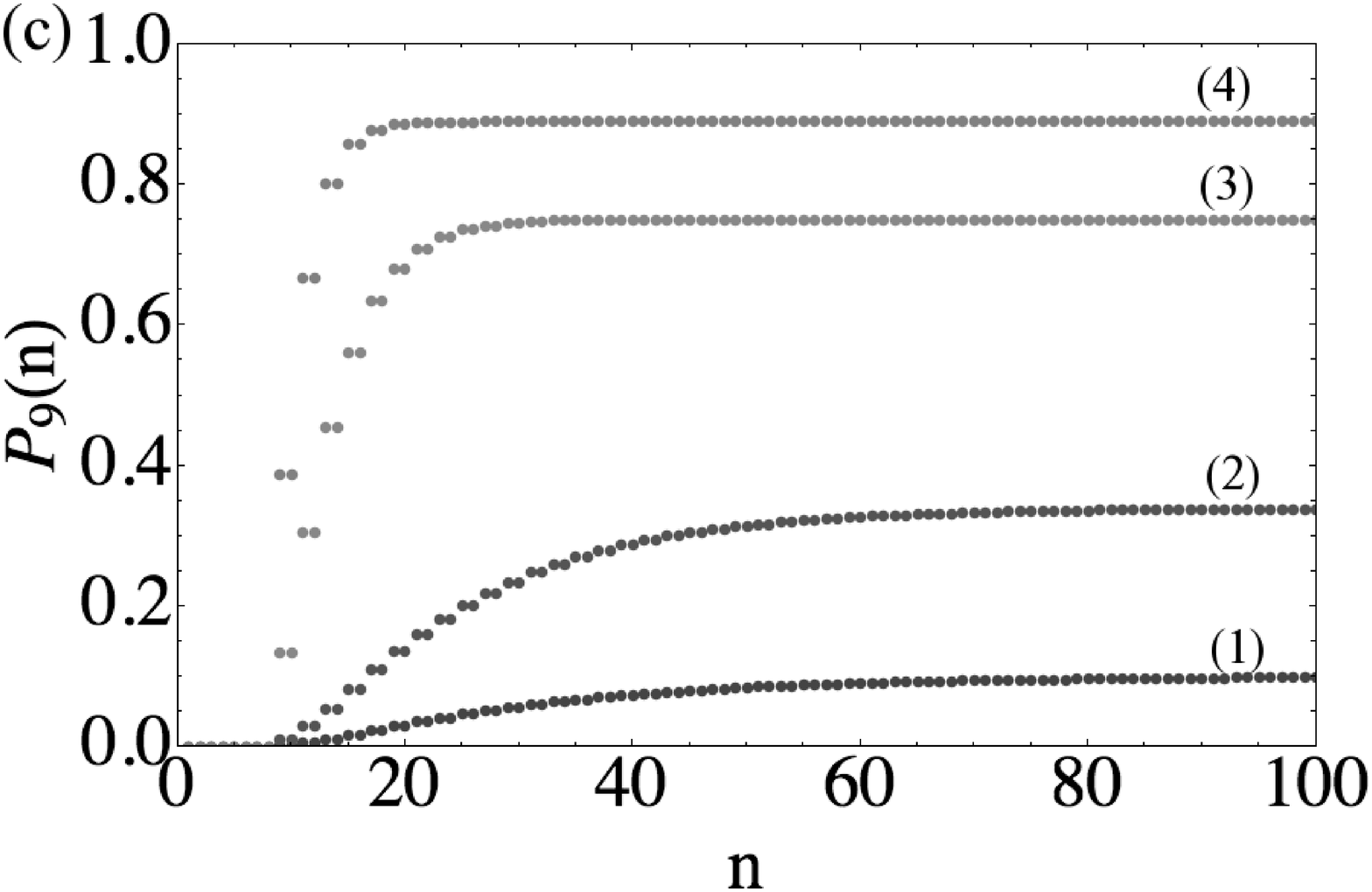}
\includegraphics[width=0.52\textwidth]{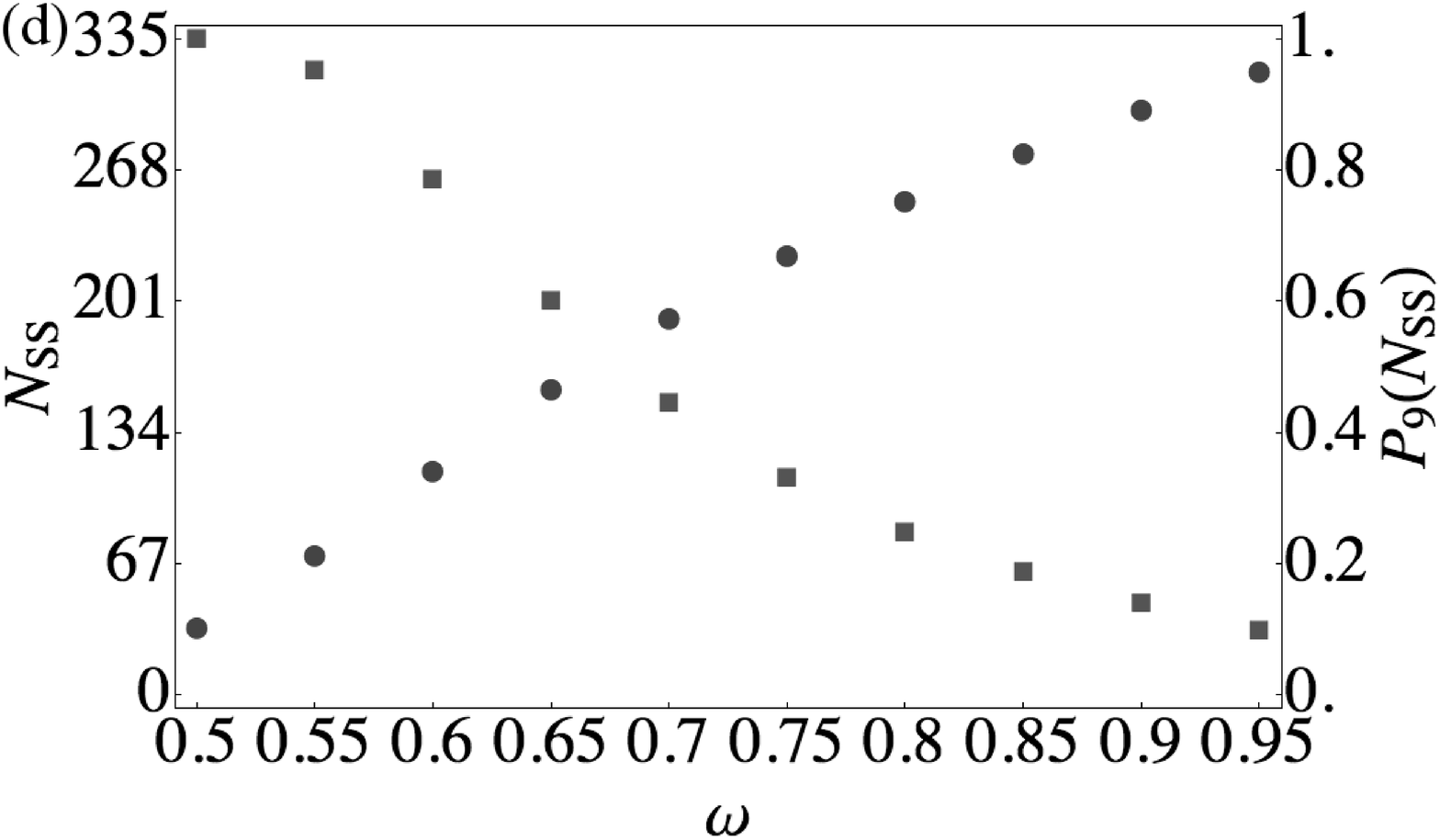}
\caption{Quantum circuit, corresponding open quantum walk diagram and efficiency of the 3-qubit QFT. Fig. 3a depicts the circuit implementation of the 3-qubit QFT. The corresponding open quantum walk diagram is shown in Fig. 3b. Fig. 3c shows the dynamics of the detection probability in the final node $9$ as function of the number of steps of the OQW. Curves (c1) to (c4) correspond to different values of the parameter $\omega=0.5, 0.6, 0.8, 0.9$, respectively. Fig. 3d shows the number of steps needed  to reach the steady state (squares) and the probability of detection of the successful implementation of the quantum algorithm (circles) as function of the parameter $\omega$. The number of steps to reach a steady states is simulated with $10^{-7}$ accuracy.}
\end{center}
\label{figU2}       
\end{figure}

\begin{figure}
\begin{center}
\includegraphics[width=0.85\textwidth]{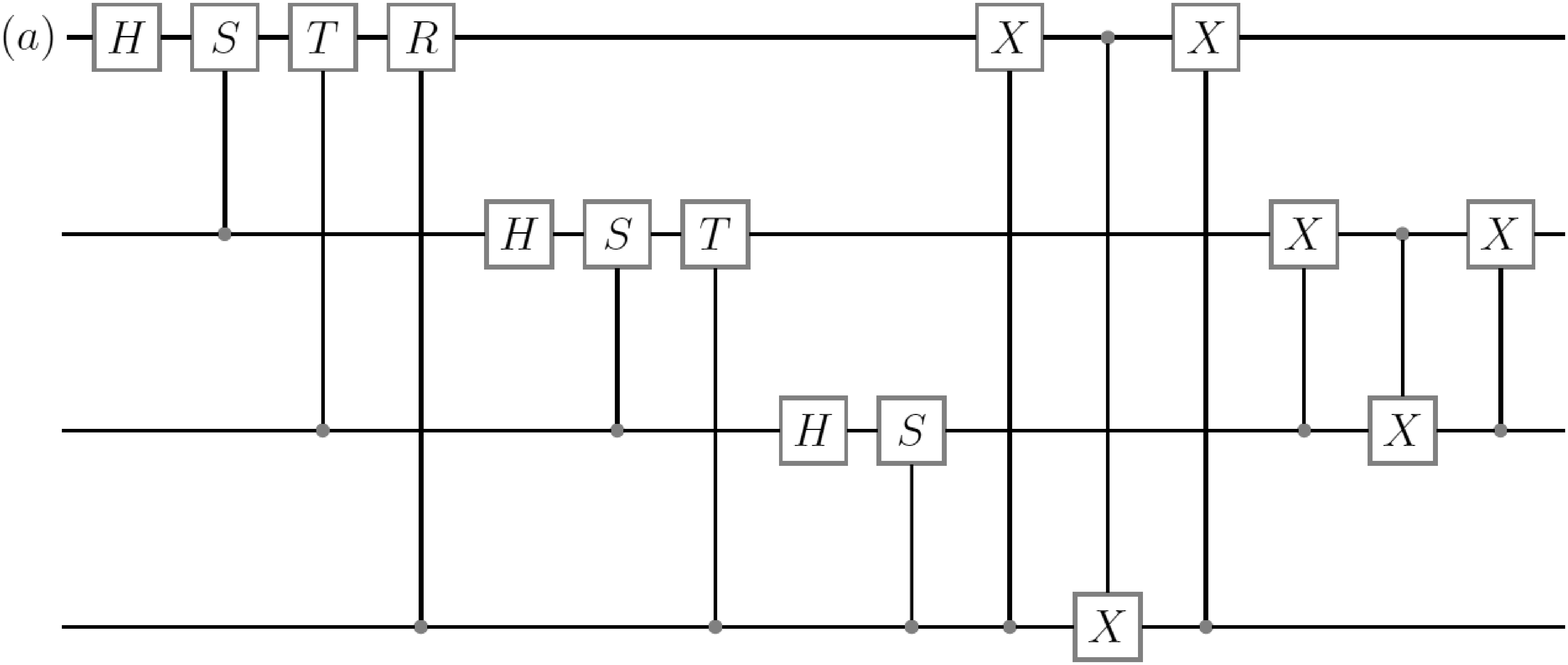}\\
\includegraphics[width=0.46\textwidth]{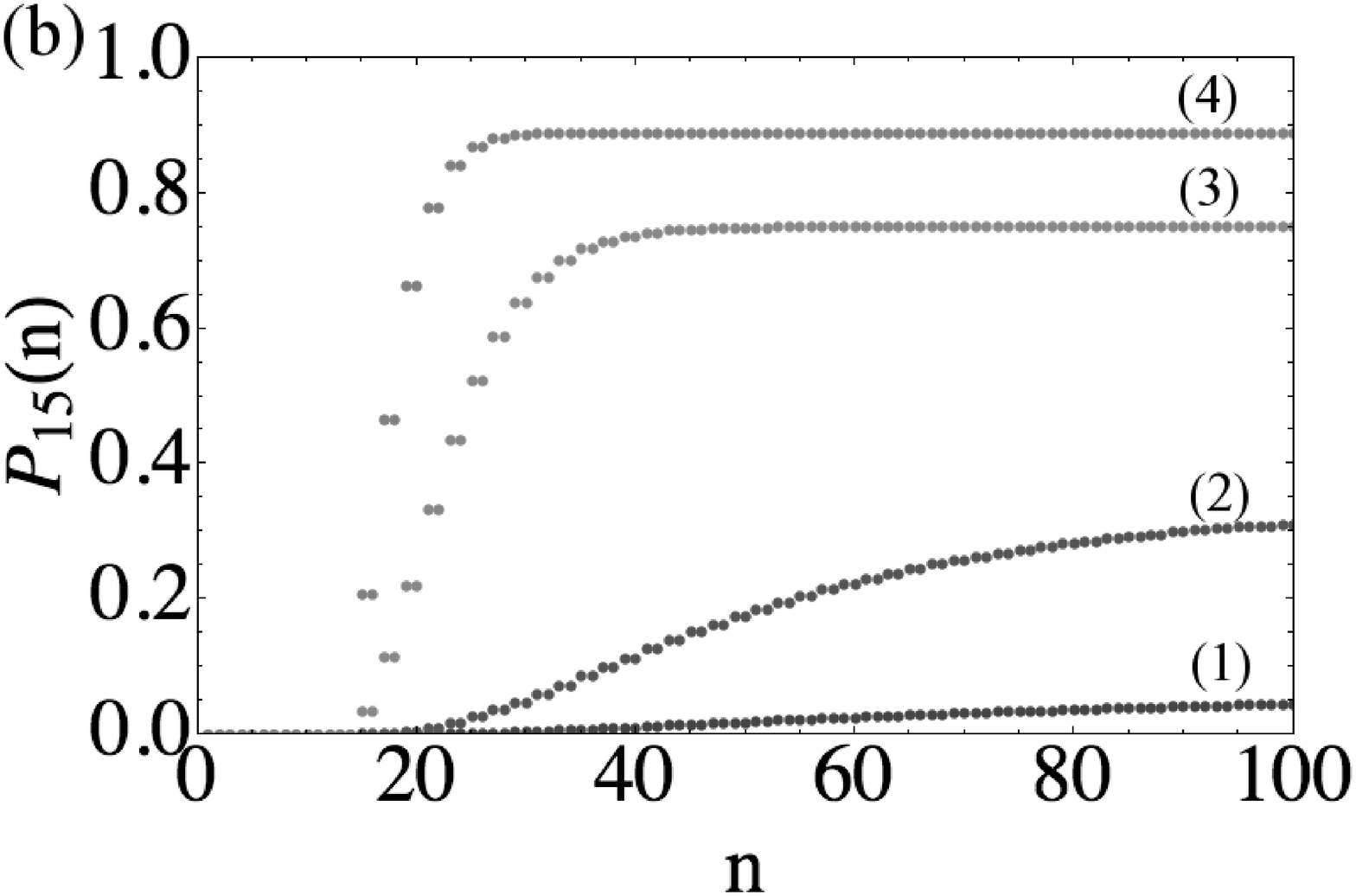}
\includegraphics[width=0.52\textwidth]{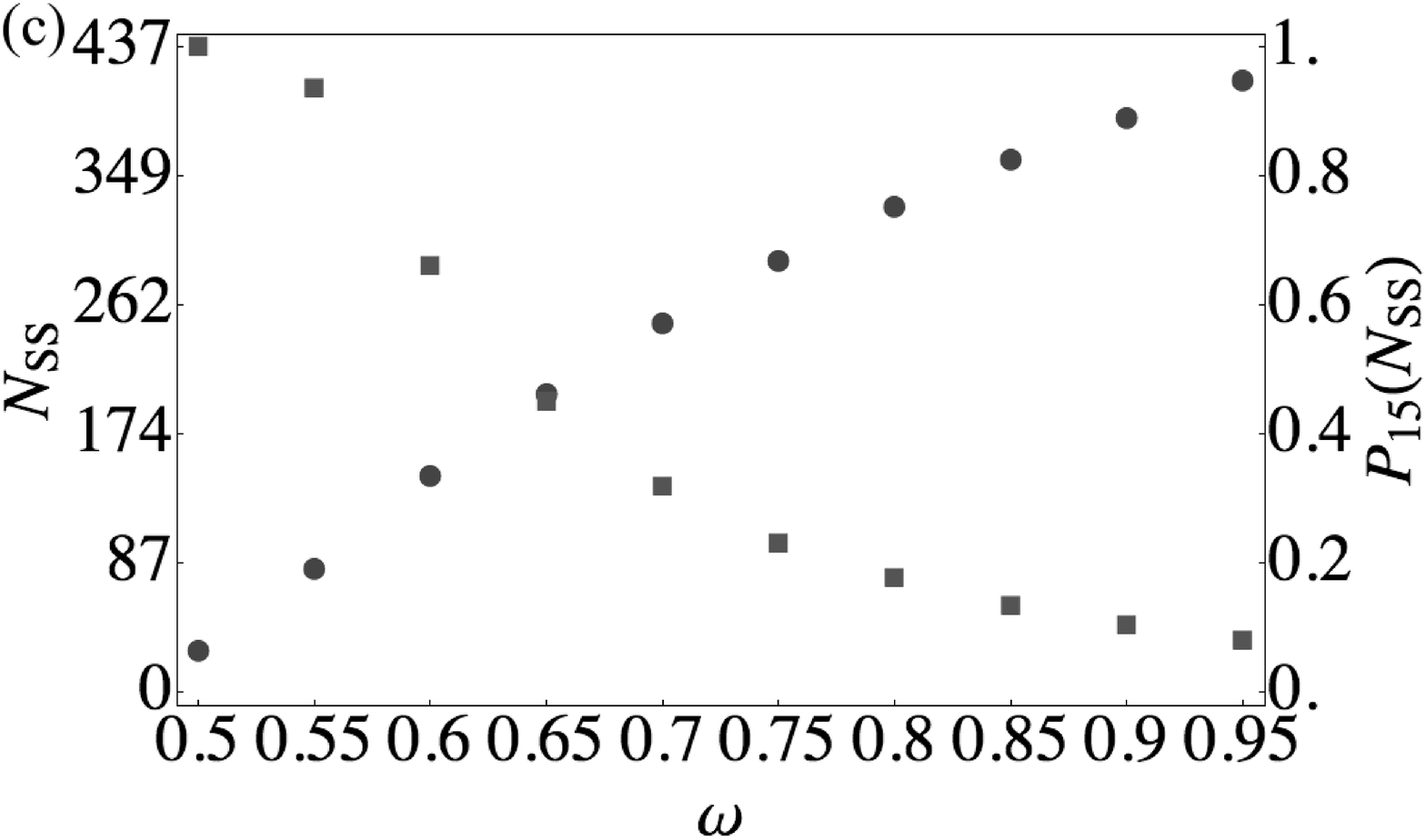}
\caption{Quantum circuit and efficiency of the 4-qubit QFT. Fig. 4a depicts the circuit implementation of the 4-qubit QFT. The corresponding open quantum walk diagram is analogous to the 3-qubit QFT (see Fig. 3b) but contains not 10 but 16 nodes. Fig. 4b shows the dynamics of the detection probability in the final node $15$ as function of the number of steps of the OQW. Curves (b1) to (b4) correspond to different values of the parameter $\omega=0.5, 0.6, 0.8, 0.9$, respectively. Fig. 4c shows the number of steps needed  to reach the steady state (squares) and the probability of detection of the successful implementation of the quantum algorithm (circles) as function of the parameter $\omega$. The number of steps to reach a steady states is simulated with $10^{-5}$ accuracy.}
\end{center}
\label{figU3}       
\end{figure}

The Quantum Fourier Transform (QFT) plays an important role in quantum computing and it is an essential part of many quantum algorithms \cite{nc}. In this section we analyze the efficiency of the OQW implementation of QFT for the example of three and four qubits. The QFT is implemented throughout a sequence of Hadamard operations, phase gates and swap-gates. The swap gates can be implemented as a sequence of three CNOT-gates for each pair of qubits. The quantum circuits for three and four qubits QFT are shown in Figs. 3a and 4a, respectively. The single qubit phase gate $R$ from Fig 4a is given by $R=|0\rangle\langle 0|+e^{i\pi/8}|1\rangle\langle 1|$. The corresponding open quantum walk diagram for a 3 qubit QFT is depicted in Fig. 3b. In the case of a 4 qubit QFT the diagram will be similar, but there will be 16 nodes. Figs. 3c and 4b show the dependence of the probability of successful performance of the QFT as a function of the number of steps of the walk. Curves (1)-(4) in both Figs. 3c and 4b correspond to different values of the parameter $\omega=0.5, 0.6, 0.8, 0.9$, respectively. As in the case of the Toffoli gate, curve (1) corresponds to the case $\omega=0.5$ which is the conventional dissipative quantum computing model. In the Figs. 3d and 4c we analyze the necessary number of steps to reach the steady state and the success probability of measurement as a function of the parameter $\omega$. Similarly to the Toffoli gate implementation we observe that with increasing $\omega$ the number of steps to reach the steady state is decreasing and the probability of successful detection is increasing. Again, this is strong evidence that the open quantum walk approach to dissipative quantum computing is a promising one.

\section{Conclusion}
After briefly reviewing the formalism of open quantum walks on graphs and of dissipative quantum computing we have demonstrated the potential of the OQW approach for dissipative quantum computing.
With the help of the Toffoli gate and the QFT we have shown that the open quantum walk approach outperforms the original dissipative quantum computing model \cite{dqc}. By increasing the probability of forward propagation in the ``time registers" in the transition operators of the open quantum walk we can increase the probability of the successful computation result detection and decrease the number of steps of the walk which is required to reach the steady state.

In future we plan to apply the open quantum walk formalism to the development of new quantum algorithms. Inspired by the successful application of  unitary quantum walks to quantum search algorithms, we expect dissipative quantum search algorithms based on open quantum walks to be an interesting alternative. Of course, the crucial milestones for the universal usage of open quantum walks for dissipative quantum computing, will be the demonstration of a physical realization procedure. The current formulation of OQWs is Markovian by design. A microscopic derivation of OQWs will assume a weak coupling of the system to the environment, so that we still can apply the standard Born-Markov approximation. Also, it will be interesting to generalize this approach to non-Markovian OQW and see if this increases further the efficiency of the implementation of dissipative quantum computing algorithms. Work along these lines is in progress.

\begin{acknowledgements}
This work is based upon research supported by the South African Research Chair Initiative of the Department of Science and Technology and National Research Foundation.
\end{acknowledgements}

\end{document}